\documentclass{aastex}
\usepackage{epsfig}
\begin{document}

\def\ni{\noindent}
\def\be{\begin{equation}}
\def\ee{\end{equation}}
\def\stl{\stackrel{<}{\sim}}
\def\stg{\stackrel{>}{\sim}}
\shorttitle{Frequency ...}
\shortauthors{J. Gil}

\title{Frequency dependence of pulsar radiation patterns}

\author{J. Gil\altaffilmark{1}, Y. Gupta\altaffilmark{2}, P.B. Gothoskar\altaffilmark{2, 3} and J. Kijak\altaffilmark{1}}
\altaffiltext{1}{Institute of Astronomy, Zielona G\'ora
University, Lubuska 2, 65-265, Zielona G\'ora, Poland}
\altaffiltext{2}{National Centre for Radio Astrophysics, TIFR, Pune University Campus, Pune 411 007, India}
\altaffiltext{3}{Present Address : Veritas Software India Pvt. Ltd., Pune 411020, India}

\email{jag@astro.ca.wsp.zgora.pl}

\begin{abstract}

We report on new results from simultaneous, dual frequency, single pulse observations 
of PSR B0329+54 using the Giant Metrewave Radio Telescope.  We find that the longitude 
separation of subpulses at two different frequencies (238 and 612 MHz) is less than 
that for the corresponding components in the average profile.  A similar behaviour 
has been noticed before in a number of pulsars. We argue that subpulses are emitted 
within narrow flux tubes of the dipolar field lines and that the mean pulsar beam 
has a conal structure.  In such a model the longitudes of profile components are 
determined by the intersection of the line of sight trajectory with the 
frequency-dependent cones of maximum mean intensity, while the longitudes of 
subpulses are determined by the intersection of the line of sight trajectory with 
subpulse-associated emission beams.  Thus, we show that the difference in the 
frequency dependence of subpulse and profile component longitudes is a natural 
property of the conal model of pulsar emission beam.  We support our conclusions 
by numerical modelling of pulsar emission, using the known parameters for this pulsar, 
which produce results that agree very well with our dual frequency observations.

\end{abstract}

\section{Introduction}

It is generally believed that the frequency dependence of pulsar
radiation patterns (subpulses, profile components etc.) reveals
the so called radius-to-frequency mapping effect : $r(\nu)\propto\nu^{-p}$,
where $r$ is the emission altitude corresponding to the frequency
$\nu$ and $p$ is a positive exponent. The coherent radio emission
from pulsars must be generated due to some instability occuring
within the stream of relativistic plasma flowing along dipolar
field lines \citep[e.g.][]{kmms96,am98,mea00}. It is commonly
assumed that the emission process is narrow-band, that is, at a
given radio emission altitude $r$ a relatively narrow band of 
frequencies, $\Delta\nu$, centered at a certain $\nu ~(\gg\Delta\nu~)$
is generated \citep{c78,c92,kg98}. As a result of the
radius-to-frequency mapping and the diverging nature of dipolar field
lines, the longitudes of details of the observed emission (such as
peaks of subpulses and profile components, profile edges etc.)
vary with frequency.  The simplest result of this is the observation
that the overall pulse width is usually braoder at lower frequencies. 
One should realize that the actual radius-to-frequency mapping can be 
revealed only if the radiation processes can be traced along approximately
the same dipolar field line (or narrow bundle of adjacent field
lines). For this, one has to locate a narrow feature in the
radiation pattern (pulse window), which is apparently correlated
in terms of intensity at different observing frequencies. The frequency
dependence of the longitude of occurrence of such correlated emission
features is given by $\varphi(\nu)\propto\rho\propto
r^{1/2}(\nu) \propto\nu^{-p/2}$, where $\rho$ is the opening angle
of dipolar field lines. The longitudinal phase
$\varphi(\nu)=360^\circ\cdot t(\nu)/P$ is determined by the time
of arrival $t(\nu)$ of a particular detail of emission pattern
within the $360^\circ$ pulse window, where $P$ is the pulsar
period. Usually the reference time of arrival ($t_0=0$) corresponds
to the fiducial plane containing the pulsar spin axis, $\bf\Omega$,
the magnetic axis, ${\bf m}$, as well as the observer, ${\bf O}$
(Fig.~\ref{fig1}). The time of arrival of radiation corresponding
to a particular feature emitted near the fiducial plane does not
depend on frequency (if the dispersive delays are properly
removed, Craft 1970). In this paper we will use the term ``detail
of emission pattern'' in several different meanings, such as: the
longitude of subpulse peak $\varphi_s$, the longitude of the mean
profile peak $\varphi_p$, the longitude of the profile midpoint
$\varphi_m$, and mean profile width $W(\nu)=2\cdot \varphi_w(\nu)$
at some intensity level $w$ referred to the maximum intensity
(usually 10\% or 50\%), where $\varphi_w(\nu)$ is the longitudinal
phase of mean profile corresponding to the  $w$ percent of
relative intensity.

\citet{i93} were the first to examine the frequency dependence of subpulse 
versus profile component variations for four pulsars: PSRs B0031$-$07, 
B0320$+$39, B1133$+$16 and B2016$+$28, which are well know for the prominent 
subpulse drift phenomenon (see Table 1). \citet{i93} used a technique 
which did not require simultaneous single pulse observations at different 
frequencies. They measured the parameter $P_2$, which is the spacing between 
two bands of drifting subpulses, for different frequencies and converted
it into a longitude separation, $\Delta\varphi_s=360^\circ\cdot P_2/P$. 
In the corresponding mean profiles they measured the pulse widths 
$\Delta\varphi_{50}=2\varphi_{50}$ and $\Delta\varphi_{10}=2\varphi_{10}$ 
at 50 and 10 per cent of the maximum intensity, respectively. \citet{i93} 
found that if one presents the frequency dependence of pulsar radiation 
patterns in the exponential form $\Delta\varphi_w\propto\nu^{-p_w}$, then 
$p_{50}>p_{10}>p_s$. In other words, they demonstrated that the longitudes 
$\varphi_s$ of subpulse peaks vary with frequency much less than the 
longitudes of corresponding profile components $\varphi_{50}$ or 
$\varphi_{10}$. From their sample of four pulsars, typcial values obtained
are $p_{50}\sim 0.35$ and $p_s\sim 0.1$ (Table 1).

As \citet{i93} noticed correctly, their results imply that the
excitations of subpulse emission do not follow exactly the
distribution of excitations leading to the mean emission, which is
difficult to reconcile with a simple pulsar picture. However,
\citet{gk96} demonstrated that this is exactly what one should 
expect if the mean pulsar beams have a conal structure. 
In fact, within a conal model of pulsar beams
\citep{r93,gks93,gk96,md99} the subpulse enhancement follows a
narrow bundle of dipolar field lines, while the mean emission
enhancement (corresponding to peaks of the mean profile
components) is distributed over a cone (or a number of nested
cones) of dipolar field lines (see Fig.~\ref{fig1} for
illustration). Thus, the frequency dependence of subpulse and
component peak longitudes should, in general, be different. On 
the other hand, for a patchy model of the mean pulsar beam 
\citep{lm88,m95}, both subpulse enhancements and those related to
the profile components follow more or less the same bundles of
dipolar field lines. Within such a model, the frequency dependence
of subpulse and component peak longitudes should be similar, which
is in conflict with available observational data.

In this paper we further discuss the problem of frequency
dependence of pulsar radiation patterns, analyzing new single
pulse data obtained for PSR B0329$+$54 simultaneously at 238 and
610 MHz, using the Giant Metrewave Radio Telescope (GMRT). We
confirm the conclusion obtained by \citet{gk96} who used averaged
data in their analysis, by showing the corresponding effects in
single pulse data.

\section{Single pulse simultaneous dual frequency data of PSR B0329$+$54}

The single pulse simultaneous dual frequency data used in our analysis 
were taken with the Giant Metrewave Radio Telescope (GMRT) at Khodad 
in India (see \cite{swarup} for details about the GMRT).  The data were 
obtained by adding the total power detected signals from 12 of the 30 
antennas of the GMRT, using the incoherent mode array combiner of the 
GMRT (see \cite{gupta} for more details about the pulsar mode of operation 
of the GMRT).  The bandwidth used was 16 MHz, divided into 256 spectral 
channels by the digital back-ends.  The raw data were integrated to a 
time resolution of 0.516 millisecond (corresponding to a pulse longitude 
resolution of 0.26 degree) before being recorded for off-line analysis, 
where the data were dedispersed and gated to obtain the single pulse 
data sequence.  A special feature of these observations was that they 
were carried out at two radio frequencies simultaneously, by splitting 
the array of 12 antennas in to two sub-arrays -- one consisting of 5 
antennas operating in the 238 MHz band and the other consisting of 7 
antennas operating in the 610 MHz band.  After total power detection, 
the data from antennas at both frequencies were added together in the 
incoherent mode array combiner.  During the off-line analysis, the 
pulsar's dispersion curve was used to discriminate between the longitude 
location of the pulses at the two radio frequencies, thus allowing the 
recovery of dedispersed pulse trains at the two different radio frequencies 
from a single raw data set.  The technique works as long as the combination 
of the two chosen frequencies, and the DM and period of the pulsar is such 
that pulses from the two radio bands (after dispersion) do not overlap.  
The advantage of the technique is that the data from the two frequencies 
is completely aligned in time, except for the dispersion delay between the 
two radio frequencies.  This delay is easily and accurately estimated and 
the two dedispersed data trains are aligned by removing this delay.  The 
data set used here was recorded on 10th August, 1999 and consists of a 
sequence of 1475 pulses of PSR B0329+54.

The average profiles for PSR B0329+54 obtained at 238 and 610 MHz from 
this data are shown in Fig.~\ref{fig2}.  Both profiles show a central 
core component and the two prominent conal components (one on each side 
of the core) that are normally easily seen for this pulsar.  The 
radius-to-frequency mapping effect for the average profiles is also clearly 
seen in this figure, with the conal components at the lower frequency 
(238 MHz) occuring further away from the central core component than 
the corresponding conal components at the higher frequency (610 MHz).
A sample of selected single pulses at the two frequencies is shown in 
Fig.~\ref{fig3}, along with the integrated profiles for reference.  The 
data for 238 MHz is shown by the thicker lines, whereas the 610 MHz 
data is shown by the thinner lines.  These single pulses are selected 
on the basis of good signal to noise ratio (at both frequencies) for 
at least one of the conal components.  The threshold used was a peak 
SNR criteria (in the longitude window for the conal component) of 7 
times the rms of the off-pulse noise.  This yielded 20 and 16 pulses 
for the leading and trailing components respectively, from which the 
5 best examples -- where the sub-pulse shapes at both frequencies are 
fairly clear and Gaussian-like -- were selected for each conal component.
The longitude windows used for the two conal components are shown in 
the figure by the grey vertical lines in the left hand panels, and are 
demarcated as ``win1'' and ``win2'' for the leading and trailing conal 
components, respectively.

It is seen that the emission features in the individual pulses also 
exhibit the radius-to-frequency mapping effect in that the subpulses 
at the lower frequency appear earlier for the leading conal component 
and later for the trailing component, when compared with the corresponding 
feature at 610 MHz. However, the interesting point is that, for all 
the single pulses shown, the longitudinal separation between the subpulse 
peaks at the two frequencies is {\it smaller} than the corresponding 
separation in the average profile. This is quantified by examining the 
peaks of the cross-correlation functions (CCFs) for the average profile 
as well as the individual pulses.  These CCFs are displayed in the
right hand panel of Fig.~\ref{fig3}. The peaks of the CCFs (marked
by the open circle symbols) are a measure of the longitudinal
separation between the correlating single pulse features at the
two frequencies, for the selected conal component.  For the
average profile, the location of the CCF peaks are marked by the
two grey vertical lines in the right hand panel.  These correspond
to a mean shift of 7 bins (1.82 degrees) of pulse phase (negative)
for the leading conal component and 4 bins (1.04 degree) of pulse
phase (positive) for the trailing conal component, for the average
profile.  These two shifts are not equal in amplitude, with the
departure from symmetry being about $(7-4)/2=1.5$ bins or 0.39
degree.  This asymmetry, which is also clearly visible to the eye 
in Fig.~\ref{fig2}, is believed to be due to the effects of
retardation and aberration \citep[see][]{gg01} which cause the
conal beams at different frequencies (emitted at different heights) 
to be aberrated and retarded by different amounts, leading to increased 
separation for the leading parts of the profile and reduced separation 
for the trailing parts of the profile. The retardation-aberration shift 
of 0.39 degree or 0.77 msec, translates into a difference of emission 
altitudes of $\Delta r\sim c\, 7.7\times 10^{-4}{\rm s}\sim 2.3\times
10^{7}$~cm, where $c=3\times 10^{10}$~cm/s. This is perfectly
consistent with the empirical radius-to-frequency mapping given by
\citet{kg98}. In fact, for PSR~B0329$+$54 \citet{kg98} give the emission 
altitude $r(\nu)=(5\pm 0.5)\times 10^{7}\nu_{GHz}^{-0.21\pm 0.07}$~cm. 
For our frequencies of 0.24 and 0.61 GHz, the emission altitudes are
about $7\times 10^{7}$~cm and $5\times 10^{7}$~cm, respectively.
Thus, the difference in altitudes is $\Delta r\sim 2\times 10^7$~cm, in
good agreement with $\Delta r\sim 2.3\times 10^7$~cm inferred
from the observed retardation-abberation shift. These numbers are
also in good agreement with the estimates for emission heights for
the conal components for this pulsar obtained by \citet{gg01}
using accurate measurements of the retardation and aberration
effects.

The typical deviation of the peak of the CCF for the individual
pulses with respect to that for the average profile for the
corresponding conal component, is 2 to 3 bins (0.52 to 0.78
degree) of longitude, with a minimum to maximum range of 0 to 5
bins (0 to 1.3 degree).  There appears to be some evidence that
these shifts in the CCF peaks are systematically larger for the
trailing conal component.  Note that there are practically no
cases where the magnitude of the shift for the individual subpulse
is more than that for the corresponding conal component in the
average profile.

The above conclusions are further quantified by studying the 
distribution of the CCF peaks for the individual pulse cases, 
compared to the peak of the CCF for the average profile for 
the corresponding conal component.  
This is illustrated in Fig.~\ref{fig4} which shows the histograms of 
the distributions of the CCF peaks obtained for the selected single 
pulses for a SNR threshold of 5 times the off-pulse rms noise.  
The results for the two conal components are shown as separate 
histograms, for the respective windows.  The total number of 
pulses that meet the threshold criteria are 56 and 57 for the 
two conal components, out of a total of 1475 pulses.  The 
actual quantity used for the histogram plots is the difference 
between the CCF peak for the individual pulse and the CCF peak 
for the average profile for the corresponding conal component.  
Thus the zero position bin in the histogram corresponds to cases 
of single pulses where the shift in the subpulse at the two 
frequencies is the same as the corresponding shift in the average 
profile.  These plots clearly show that the CCF peaks for the 
individual pulses are asymmetrically distributed with respect 
to the CCF peak for the average profile for the corresponding 
conal component. They also show that the magnitude of the shift 
is somewhat larger for the trailing component, as compared to the 
leading component, with mean values for the shifts being -2.5 bins 
(-0.65 deg) and 1.4 bins (0.36 degree), respectively.

These results are consistent with those of \citet{i93}. All these 
are summarized in Table 1. Although there are differences from
pulsar to pulsar, the results in Table 1 reinforce the conclusion
that the frequency dependence of subpulse patterns is apparently
weaker than the frequency dependence of corresponding features in
the average profile. The exponents quoted for PSR B0329+54 (0.14
and 0.07 for mean profile and single pulses, respectively) refer
to average values following from estimates obtained for the left
(L) and right (R) side of the pulse profiles, after corrections
for retardation-aberation shift of about 1.5 bin (0.4 degree of
longitude). In the next section we argue that this different
frequency dependence of subpulses and corresponding profile
components is natural, provided that the mean pulsar beam has a
conal structure and the subpulse emission is associated with
narrow bundles of dipolar magnetic field lines. We assume that
only one bundle (controlling the subpulse-associated plasma flow)
is related with one subpulse observed simultaneously at the two
frequencies.

\section{Structure of pulsar beams}

It is widely believed that pulsar elementary emission is
relativistically beamed tangently to dipolar magnetic field lines.
Thus, the geometry of the emission region can be described by the
opening angle $\rho(\nu)=1^\circ.25~s\ r_6(\nu)^{1/2}P^{-1/2}$
(Fig.~\ref{fig1}), where $P$ is the pulsar period, and
$r_6=r(\nu)/R$ is the frequency dependent emission altitude in
units of stellar radius $R=10^6$~cm. The mapping parameter ($0\leq
s\leq 1$) is determined by the locus of dipolar field lines on the
polar cap ($s=0$ at the pole and $s=1$ at the polar cap edge).
Thus, $s=d/r_p$, where $d$ is the distance from the magnetic axis,
$\bf{m}$, to a field line on the polar cap corresponding to a
certain detail of the emission pattern, and $r_p=1.4\cdot
10^4P^{-1/2}$~cm is the canonical polar cap radius. According 
to the widely accepted concept of radius-to-frequency mapping, 
the higher frequencies are emitted at lower altitudes than the 
lower frequencies. \citet{kg98} have found an empirical formula
describing an altitude of the emission region corresponding to a
frequency $\nu_{\rm GHz}$ (in GHz) as 
\begin{equation}
r_6\approx 50\cdot\nu_{\rm GHz}^{-0.21}\cdot\tau_6^{-0.07}\cdot P^{0.33}
\end{equation}
where $\tau_6$ is the pulsar characteristic age in million years 
and $P$ is the pulsar period. This empirical relationship will be 
used in our model calculations.

To perform geometrical calculations of the radiation pattern one
has to adopt a model of instantaneous energy distribution on the
polar cap. We will assume that at any instant the polar cap is
populated by a number of features with a characteristic dimension
${\cal D}$ \citep[e.g.][]{gs00}, each delivering to the
magnetosphere a column of plasma flowing along separate bundle of
dipolar magnetic field lines. These plasma columns are supposed to
be sources of the narrow-band radiation, with different radio
frequencies emitted at different altitudes, as described by the 
radius-to-frequency mapping law \citep{kg98}. Thus, one plasma 
column illuminates different spots at different frequencies.
Some of these spots can be intersected by the line of sight (LOS henceforth) and thus
observed simultaneously at two frequencies. Each feature is
modelled at the polar cap by a gaussian intensity distribution
\begin {equation}
I_i=\exp(-\kappa l^2/{\cal D}^2)
\end{equation}
where 
$l^2=d^2(\varphi)+d_i^2-2d(\varphi)d_i\cdot\cos[\sigma(\varphi)-\sigma_i]$.
Here the pairs $[d_i, \sigma_i]$ and $[d(\varphi),
\sigma(\varphi)]$ are polar coordinates of the i-th feature
(maximum intensity) and the observer's position on the
line of sight (projected onto the polar cap), respectively. The
instantaneous emission of the $n$-th pulse is described by
$I_n(\varphi)=\Sigma_{i=1}^{K_n}I_i(\varphi)$, where $K_n$ is the
number of adjacent features contributing to the radiation observed
in the $n$-th pulse at a longitude $\varphi$. The average pulse
profile is therefore
$I(\varphi)=\frac{1}{N}\Sigma_{n=1}^{N}I_n(\varphi)$, where $N$ is
the number of averaged single pulses. The azimuthal angle variations 
of the $i$-th feature are described by 
\begin{equation}
\sigma_i={\rm arctan}[\sin\varphi_i\cdot\sin(\alpha+\beta)\cdot\sin\alpha/\cos(\alpha+\beta)-\cos\alpha
\cdot\cos\rho_i]-n\cdot\Delta\sigma ~~, 
\end{equation}
where 
\begin{equation}
\rho=2\cdot{\rm asin}[\sin^2(\varphi/2)\cdot\sin\alpha\cdot\sin(\alpha+\beta)+
\sin^2(\beta/2)]^{1/2}=1.^\circ 25\ s\ r_6(\nu)^{1/2}P^{-1/2} ~~,
\end{equation}
and $\Delta\sigma$ is the change of azimuthal angle per one
pulsar period, $\alpha$ is the inclination angle between the
rotation and magnetic axes and $\beta$ is the impact angle -- the 
angle of the closest approach of the line of sight to the magnetic 
axis (Figs.~\ref{fig1}, \ref{fig5} and \ref{fig6}).

Any specific intensity distribution can be transferred along
dipolar field lines to the emission region and then along straight
lines to a given observer specified by inclination and impact
angles ($\alpha,~\beta$). Thus, knowing a distribution of opening
angles $\rho_i(\nu)=1.^\circ 25(d_i/r_p)r_6^{1/2}(\nu)P^{-1/2}$
corresponding to instantenous radiation, one can calculate the
longitude of an $i$-th detail of the emission pattern
$\varphi_i=2\ {\rm asin}\left\{\left[\sin\left(\frac{\rho_i-\beta}{2}\right)\cdot\sin\left(\frac{\rho_i+\beta}{2}\right)\right]/
[\sin\alpha\cdot\sin(\alpha+\beta)]\right\}^{1/2}$ at a given
frequency $\nu$ \citep{g81}. The mean profile can be obtained by a
simple averaging of individual waveforms, which can in principle
vary in time, depending on the adopted model of distribution of
instantaneous arrangement of energy on the polar cap.

\citet{gk96} and \citet{gs00} argued that the nested cone
structure of mean pulsar beams is induced by the vacuum gap ${\bf
E}\times{\bf B}$ drift \citep{rs75}, which forces the
subpulse-associated plasma columns to rotate around the magnetic
axis. This model suggests a simple and natural relationship
between a sequence of single pulses and features of mean emission.
It is also very convenient for geometrical simulations, because
the azimuthal angle $\sigma$ (eq. 3) of each plasma column varies
regularly with time (sequential pulse number). The frequency
dependence of pulsar emission patterns within such a model is 
qualitatively illustrated in Fig.~\ref{fig1}. The subpulse-associated 
plasma columns corresponding to the half power beam width of the
subpulse emission (thin small circles marked by $S(\nu)$), which 
will be called spots, perform a circumferential motion around the 
magnetic axis $\bf{m}$, in which the azimuthal angle $\sigma$ varies 
with time while the opening angle $\rho(\nu)$ remains constant. This
motion determines the cones (thick large circles) of the maximum
average intensity with the opening angles $\rho_1$ and $\rho_2$ at
frequencies $\nu_1$ and $\nu_2$, respectively. The frequency
dependent longitude $\varphi_p(\nu)$ of the profile component is
determined by the intersection of the line of sight with the
average cone at the frequency dependent opening angle $\rho(\nu)$.
The frequency dependent longitude $\varphi_s(\nu)$ of subpulse
peak is determined by the local maximum intensity along the cut of
the line of sight through the subpulse spot. These cuts are marked
by dashed lines (shorter dash for $S^\prime(\nu_1)$ and longer
dash for $S(\nu_2)$, respectively). The position of the spot $S$
was choosen in such a way that
$\varphi_s(\nu_1)=\varphi_p(\nu_1)$. This is however a special
case, which does not hold for the spot $S^\prime$, in which case
the longitude of the subpulse peak $\varphi_s^\prime(\nu_1)$ is
different from the longitude of a corresponding component
$\varphi_p^\prime(\nu_1)$. When the frequency changes from $\nu_1$
to $\nu_2$, the subpulse peak longitude changes from
$\varphi_s(\nu_1)$ to $\varphi_s(\nu_2)$, while the corresponding
profile peak longitude changes from $\varphi_p(\nu_1)$ to
$\varphi_p(\nu_2)$. Note that
$\Delta\varphi_p=\varphi_p(\nu_2)-\varphi_p(\nu_1)$ is generally
larger than $\Delta\varphi_s=\varphi_s(\nu_2)-\varphi_s(\nu_1)$.
This is consistent with observations of four pulsars published by
\citet{i93} and confirmed in this paper for yet another pulsar PSR
B0329$+$54 (Table 1).

It is important to note that in this model, 
$\Delta\varphi_s\leq\Delta\varphi_p$, is {\it always} true. 
The equality condition holds only for two special cases: 
(i) $\sigma\approx 90^\circ$ and $\beta\approx 0^\circ$ and 
(ii) $\sigma\approx 0^\circ$. The latter case is trivial since it 
corresponds to $\Delta\varphi_s=\Delta\varphi_p\approx 0^\circ$ 
for which both subpulses are observed at the fiducial phase; while 
the former one corresponds to central cut of the LOS 
across the beam. For all cases other than these two, 
$\Delta\varphi_s<\Delta\varphi_p$ is strictly true.
This prediction of the model has been shown to be true in the 
observed data (Figs.~2 and 3) and will be shown to hold for the 
simulated synthetic data as well. 
It should be emphasized that we consider only cases in which both higher 
and lower frequency spots illuminated by the same single 
subpulse-associated plasma column are observed (i.e. intersected by 
the LOS). The cases in which subpulse at one of the two frequencies 
is missing are not interesting for the present analysis, as no 
frequency dependent effect can be measured for these. 

After qualitative illustration of the problem (Fig.~\ref{fig1}) we
now explore it quantitatively by means of geometrical
simulations. Although we did not intend to model exactly the
pulse profiles of PSR B0329$+$54, we adopted in our calculation
the values of $P=0.7$~s and $\dot{P}=2\cdot 10^{-15}$. Thus, our
model of the polar cap or the corresponding instantaneous
radiation beam presented in Fig.~\ref{fig5} corresponds in a sense
to PSR B0329$+$54. We followed \citet{gs00} (their eqs.~[11] and [12])
for description of the polar cap arrangement.  The
characteristics of instantaneous radiation are imprinted on the
polar cap, transmitted along dipolar field lines to the radio
emission altitudes and then emitted to a particular observer
along straight paths. The circumferential motion
of sparks around the magnetic axis (Fig.~\ref{fig5}) will lead to
a nested cone structure of the average radiation beam, as
presented in Fig.~\ref{fig6}. The adopted inclination angle
$\alpha=30^\circ$ and four line of sight trajectories correspond
to impact angles $\beta=0,~3.3,~5.0$ and 6.6 degrees,
respectively.  We performed calculations for the two frequencies
238 and 610 MHz. Hence, each line of sight trajectory bifurcates
when projected onto the surface of the polar cap.

The results of calculations of radiation waveforms corresponding 
to emission patterns presented in Figs.~\ref{fig5} and \ref{fig6} 
are shown in Fig.~\ref{fig7}. Four vertical panels correspond to 
different impact angles $\beta=0,~3.3,~5.0$ and 6.6 degrees, 
respectively. The broader emission at 238 MHz is shown by dashed 
lines and higher frequency radiation at 610 MHz is shown by solid 
lines.  Both instantaneous single pulse and mean profile are 
presented for each impact angle. We have calculated a long sequence 
of single pulses to make sure the mean profile is stable. We selected 
pulse \#4 in our sequence (for no particular reason) and show its 
profile for each impact angle at the two frequencies.

The first panel corresponds to a central cut of the line of sight 
with the impact angle $=0^\circ$ and, of course, $\sigma\sim
90^\circ$.  In this case the longitudinal
difference of subpulse and profile component peaks occuring at
different frequencies is exactly the same
$(\Delta\varphi_s^{max}=\Delta\varphi_p)$. This can be seen from
both visual inspection of the pulse profiles and from the CCF
analysis. The CCF computed separately for left (L) and right (R)
component, peaks at the same lag in both mean and single pulses.
However, as $\beta$ increases, the difference for subpulses becomes 
smaller and smaller as compared with a corresponding difference for 
peaks of profile components. This translates into CCF lags which 
are smaller for single pulses than for mean pulses. The lag difference 
is about 0.5, 1.0 and 2.0 degrees for $\beta$ values of 3.3, 5.0 
and 6.6 degrees, respectively.  This can be compared with Fig.~\ref{fig3}, 
where the mean lag difference is about 2 sample bins or 0.5 degrees 
of longitude. Thus, the actual data presented in Fig.~\ref{fig3} 
correspond to a modelling case with $\beta\sim3$ degrees (see Fig.~7).

Another interesting, though obvious property is that the frequency 
separation disappears near the fiducial longitude $\varphi=0^\circ$, 
as in the actual data presented in Figs. ~\ref{fig2} and ~\ref{fig3}.
We have selected three central subpulses marked by 1, 2 and 3 in 
Fig.~\ref{fig7}.  The CCF computed for these subpulses show lags 
near zero in Fig.~\ref{fig8} (for the subpulse 1 the lag is 
exactly zero).  The explanation of this effect is simple and 
natural. The fiducial logitude was chosen in such a way that it 
corresponds to the fiducial plane containing both the $\bf\Omega$ 
and $\bf{m}$ axes as well as the observer $\bf{O}$. The divergence 
of dipolar magnetic field lines does not affect longitudinal phases 
of the observed radiation emitted exactly at this plane (subpulse 1) 
and affects it only slightly near this plane (subpulses 2 and 3).

In our simulations we have not included the effects of retardation
and aberation, which were unequivocally detected in our high
resolution, simultaneous dual frequency single pulse data. These
effects shift the entire low frequency emission patterns towards
earlier phases (leading side) with respect to high frequency
patterns. This results in larger frequency difference on the left
(L) than on the right (R) side of the profile, thus producing the
asymmetrical CCF lags seen in Fig.~\ref{fig3}. However, the
deviation of single pulse lags from mean pulse lags does not
depend on the retardation-aberation shift (see Fig.~\ref{fig2}),
so our symmetrical model CCF can be compared with the asymmetrical
CCF computed for the actual data.

\section{Discussion and conclusions}

The observed emission patterns of pulsar radio emission are
usually frequency dependent. The longitudes of details of pulse
profiles appear earlier/later for the leading/trailing side of the
pulse window, as frequency decreases. This is well understood in
terms of the radius-to-frequency mapping and diverging nature of
dipolar magnetic field lines in the radio emission region.
However, in PSR B0329+54 and four other pulsars (Table 1) it has
been observed that for emission at two different frequencies, the 
difference in longitudes of subpulse peaks is generally smaller than 
the difference in longitudes of corresponding component peaks. This
suggests that the subpulse and profile component excitations do
not follow the same bundles of dipolar field lines, which appears
surprising at first sight. However, we argued in this paper that
this is exactly what one should expect if:
\begin{description}
\item (i) the subpulse enhancements are related to relativistic plasma 
columns following a narrow (and separated from each other) bundle 
of dipolar field lines, with different frequencies generated at 
different altitudes $r_6(\nu)=r(\nu)/R$, 
\item (ii) the profile component enhancements are related to the 
conal structure of the mean pulsar beam, and
\item (iii) the line of sight trajectory of the observer across 
the beam is neither central not grazing (i.e. the impact angles 
$0<\beta<\rho_b$, where $\rho_b=1^\circ.25\ r_6^{1/2}P^{-1/2}$ is 
the opening angle of the full pulsar beam). Particularly, for 
PSR B0329+54 we deduced the impact angle $\beta\approx 3^\circ$, 
which agrees well with other independent estimates from about 
$2^\circ$ \citep{r93} to about $4^\circ$ \citep{gl95}.
\end{description}

In deriving our conclusions, we have made a number of well 
justified assumptions, which we summarize below:

\begin{description}

\item (1) The pulsar radiation is relativistically beamed along 
dipolar lines of a global magnetic field of the neutron star.  
This is a generally well accepted assumption.

\item (2) The narrow-band pulsar radio emission follows the 
radius-to-frequency mapping $r=r(\nu)$. In our calculations
we used a specific form of the empirical relation obtained by
\citet{kg98}. Although, the actual form of the radius-to-frequency
mapping is not important for our general conclusions (as long as
the higher frequencies are generated closer to the polar cap than
the lower ones), this relation is strongly supported by the
retardation-aberation shift clearly revealed in our
high-resolution, simultaneous dual frequency data.

\item (3) The instantaneous emission beam consist of a number 
of subpulse-associated beams emitted by the spark-associated 
plasma columns. Thus, we invoke a non-stationary model of inner 
accelerator discharge via a number of isolated sparks \citep{s71,rs75,gm01}.


\item (4) The structure of mean pulsar beam results from a 
behaviour of subpulse-associated beams averaged over time
intervals much longer than the pulsar period. In our model 
these beams perform a circumferential motion around the 
magnetic axis following the  ${\bf E}\times{\bf B}$ drift 
motion of sparks on the polar cap.  Strong observational evidence 
of such circumferential motions has been recently presented 
by \citet{dr99,dr01}.

\end{description}

The model of the polar cap arrangement we have adopted here is
highly symmetric, which simplifies geometrical calculations.
However, our general conclusions do not depend on any specific
details of this model, as long as the instantaneous beam is
organized into a group of isolated subbeams (Fig.~\ref{fig5}) and
the average beam has a conal structure (Fig.~\ref{fig6}). In our
model this conal structure is caused by the circumferential motion
of spark-associated, subpulse-producing plasma columns around the
axis of symmetry (magnetic axis). For a convenience, we  have
calculated the emission patterns using a specific velocity of
circumferential motion following from the ${\bf E}\times{\bf B}$
drift applied to sparks (eq.~[14] in Gil \& Sendyk 2000), although
the value of this velocity is not critical for our results. Our
general conclusions apply to a more general picture of the
organization of pulsar beams, in which enhancement of excitations
leading to subpulses and corresponding profile component,
respectively, do not follow each other. We consider the ${\bf
E}\times{\bf B}$ drift induced model as the simplest and most
natural, but the reason for the formation of the conal structure
of the mean pulsar beam is not important for our conclusions. We
would also like to emphasize that our conclusions do depend
sensitively on the impact angles $\beta$ of the closest approach
of the line of sight to the magnetic axis ${\bf m}$
(Fig.~\ref{fig1}) but they are rather independent of the
inclination angles $\alpha$.

To conclude, we have demonstrated in this paper that the angular
beaming model of instantaneous emission accompanied by the conal
model of mean emission predict frequency dependent properties of
pulsar radiation that match well with those observed in the
simultaneous, dual frequency single pulse data of PSR B0329$+$54
that is reported here.  This model also explains the earlier
observations reported for four other pulsars -- PSRs B0031$-$07,
B0320$+$39, B1133$+$16 and B2016$+$28.  We therefore conclude
that, at least in these pulsars, the mean pulsar beams have a
conal structure, possibly induced by the circumferential ${\bf
E}\times{\bf B}$ drift motion of subpulse-associated beams. Thus,
our paper provides independent support for the nested-cone model
of pulsar beams \citep{r93,gks93,kwj94,md99,gg01}.

\begin{acknowledgements}
This paper is supported in part by the Grant 2~P03D~008~19 of the
Polish State Committee for Scientific Research. We thank an
anonymous referee for useful comments. We are thankful to Drs. D.
Mitra and B. Zhang for helpful discussions. We thank E. Gil, O.
Maron and M. Sendyk for technical assistance.  We also thank the
staff of the GMRT that made these observations possible and
specifically Dr. B.C. Joshi for this particular data set. GMRT is
run by the National Centre for Radio Astrophysics of the Tata
Institute of Fundamental Research.
\end{acknowledgements}

\clearpage
\begin{deluxetable}{cccccc}
\footnotesize
\tablecaption{Frequency dependence of pulse profile characteristics $\varphi_i\propto\nu^{-p_i}$.\label{tbl1}}
\tablewidth{0pt}
\tablehead{
\colhead{PSR~B} & \colhead{$\varphi_{50}\propto\nu^{-p_{50}}$} & \colhead{$\varphi_{10}\propto\nu^{-p_{10}}$} &
\colhead{$\varphi_{c}\propto\nu^{-p_{c}}$}   & \colhead{$\varphi_{s}\propto\nu^{-p_{s}}$} & \colhead{Ref.}}
\startdata
0329$+$54 &      &      & 0.14 & 0.07 & (2)\\
0031$-$07 & 0.45 & 0.35 & & 0.05 & (1) \\
0320$+$39 & 0.34 & 0.18 &  & 0.15 & (1)\\
1133$+$16 &      &      & 0.25 & 0.06 & (1)\\
2016$+$28 & 0.35 & 0.2 &  & 0.17 & (1)\\
\enddata
\tablecomments{$\varphi_{50}$ - half width\ $W_{50}=2\cdot\varphi_{50}$ of the profile at 50\% maximum intensity,
$\varphi_{10}$ - half width $W_{10}=2\cdot\varphi_{10}$ of the profile at 10\% maximum intensity,
$\varphi_c$ - longitude of profile component peak,
$\varphi_s$ - longitude of subpulse peak,
Ref: (1) \citet{i93}, (2) this paper }

\end{deluxetable}

\clearpage
\newpage
\centerline{\large\bf Figure captions}

\bigskip

\figcaption[]{The conal model for the mean pulsar beam.  The geometry of observation is 
determined by the inclination angle $\alpha$ between the magnetic axis $\bf{m}$ and the 
spin axis $\bf\Omega$ and the impact angle $\beta$ of the closest approach of the line 
of sight (observer ${\bf O}$) to the magnetic axis $\bf{m}$. The plane containing the 
spin axis $\bf\Omega$, the magnetic axis $\bf{m}$ and the observer $\bf{O}$ defines the 
fiducial longitude $\varphi=0^\circ$.  The frequency dependent position of any observed 
feature in the beam pattern is described by two angles: the frequency dependent opening 
angle $\rho(\nu)$ between the ${\bf m}$ axis and the line of sight and by the azimuthal 
angle $\sigma$ between the fiducial plane and the plane of dipolar field lines associated 
with a particular feature. The subpulse emission is associated with the frequency 
dependent subpulse spots $S(\nu)$ and the profile components are associated with the 
frequency dependent cones. The longitudes of subpulse peaks $\varphi_s(\nu)$ and profile 
peaks $\varphi_p(\nu)$ are marked, with $\nu_2>\nu_1$.
\label{fig1}}

\figcaption[]{Average profiles at 238 and 610 MHz for PSR B0329$+$54, obtained from 
the simultaneous dual frequency observations.  The darker curve is for 238 MHz and the 
lighter one for 610 MHz. Each sample bin corresponds to 0.26 degrees of longitude.
\label{fig2}}

\figcaption[]{The observed frequency dependence of radio emission patterns in PSR B0329$+$54.  
Selection of 10 single pulses from the observations is shown (at both frequencies) in the 
left hand panels.   The darker curves are for 238 MHz and the lighter curves for 610 MHz.
The average profile (top panel) is shown truncated in amplitude to highlight the conal 
components.  The right hand panels show the cross-correlation functions for the relevant 
conal component, with the peaks marked by the open circle symbols. Each bin or lag 
corresponds to 0.26 degrees of longitude.
\label{fig3}}

\figcaption[]{Histograms of the distributions of the location of the CCF peaks obtained 
for the leading conal component (top panel) and trailing conal component (bottom panel) 
for the selected single pulses.  The locations are taken with respect to the location of 
the peak of the CCF for the average profile for the corresponding conal component.
\label{fig4}}

\figcaption[]{Instantaneous arrangement of sparks on the Polar Cap (or instantaneous pulsar 
beam projected onto the PC along dipolar field lines) corresponding to the simulated pulse 
\#4 presented in Fig.~\ref{fig7}. Four different (frequency bifurcated) line of sight 
trajectories with impact angles $\beta=0.0$, 3.3, 5.0, 6.6 degrees corresponding to 
frequencies 238 and 630 MHz are presented. The pulsar rotates clock-wise, so the leading 
part of the pulse corresponds to the right side of this figure (or left if viewed from the 
spin axis $\bf\Omega$). The inclination angle between the magnetic axis (marked by the central 
dot) and the spin axis is 30 degrees.
\label{fig5}}

\figcaption[]{The average structure of the polar cap (or mean pulsar beam projected onto 
the PC) resulting from the circumferential motion of sparks (or subpulse beams) around 
the magnetic axis (marked by the central dot). For explanation of different line of sight 
trajectories see caption for Fig.~\ref{fig5}.
\label{fig6}}

\figcaption[]{The frequency dependence of pulsar radiation patterns corresponding to 
Figs.~\ref{fig5} and \ref{fig6}. In the analogy to Fig.~3, the left panels present the 
pulses, while the right panels present the corresponding cross-correlation functions (CCF). 
Each vertical panel presents the integrated profile and a single pulse for a given impact 
angle $\beta=0.0$, 3.3, 5.0 and 6.6 degrees. The symbols L and R correspond to the leading 
and trailing conal components of the profile, respectively. One can easily notice from 
both visual inspection and from the CCF analysis that the difference in longitudes of 
pulse features occuring at 238 and 610 MHz depends on the impact angle $\beta$ (closest 
approach of the observer to the beam axis). For $\beta=0$, this separation is the same 
for both single pulses and integrated profiles. However, at larger impact angles the
separation for subpulses is smaller than for a corresponding profile components.
It is worth noting that longitudes of subpulse peaks and corresponding profile components 
are the same near the fiducial phase $\varphi=0^\circ$ like in the real data presented in 
Figs.  2 and 3 (see CCF analysis in Fig.~\ref{fig8} corresponding to central subpulses 
marked by 1, 2 and 3 in this figure).
\label{fig7}}

\figcaption[]{The cross-correlation function (CCF) calculated for
three subpulses marked by 1, 2 and 3 in Fig.~\ref{fig7}. The
maximum of each function is at zero lag, meaning that subpulses
near the fiducial longitude $0^\circ$ (profile centre) appear in
the same phases at both frequencies, unlike the side subpulses
analysed in Fig.~\ref{fig7}. \label{fig8}}

\end{document}